\documentclass[fleqn,usenatbib,twocolumn]{mnras}
\usepackage[utf8]{inputenc}
\usepackage{graphicx}
\usepackage{caption,subcaption}
\usepackage{amsmath}
\usepackage{siunitx}
\usepackage{comment}
\usepackage{mathtools}
\usepackage{lipsum}

\title[Simulations of intensity interferometry]{Radius measurement in binary stars: simulations of intensity interferometry}

\author[Rai et al.]{Km Nitu Rai,$^{1}$\thanks{E-mail: niturai20129617@iisertvm.ac.in}  Soumen Basak,$^{1}$
  and Prasenjit Saha$^{2}$\\ \\
$^{1}$School of Physics, Indian Institute of Science Education and Research Thiruvananthapuram, Maruthamala PO, Vithura,\\ Thiruvananthapuram 695551, Kerala, India \\
$^{2}$Physik-Institut, University of Zurich, Winterthurerstrasse 190, CH-8057 Zurich, Switzerland}

\date{}

\begin{document}

\maketitle

\begin{abstract}
Mass and radius measurements of stars are important inputs for models
of stellar structure. Binary stars are of particular interest in this
regard, because astrometry and spectroscopy of a binary together
provide the masses of both stars as well as the distance to the
system, while interferometry can both improve the astrometry and
measure the radii of the stars.  In this work we simulate parameter
recovery from intensity interferometry, especially the challenge of
disentangling the radii of two stars from their combined
interferometric signal.  Two approaches are considered: separation of
the visibility contributions of each star with the help of differing
brightness ratios at different wavelengths, and direct fitting of the
intensity correlation to a multi-parameter model.  Full image
reconstructions is not attempted.  Measurement of angular radii,
angular separation and first-order limb-darkening appears readily
achievable for bright binary stars with current instrumentation.
\end{abstract}

\section{Introduction}

The mass and radii of stars are nowadays most often determined from
their asteroseismic frequencies, using scaling relations
\citep[e.g.,][]{1995A&A...293...87K} that relate stellar parameters to
the solar values, or by comparing the frequencies with
stellar-structure models \citep[e.g.,][]{2017MNRAS.467.1433R}.  As the
asteroseismic methods are indirect, it is desirable to compare them,
wherever feasible, with dynamical mass measurements and geometrical
radius measurements, which directs attention to binary stars.

It has been known since \cite{1928AJ.....38...89R} that combining
astrometric and photometric observations of a binary stellar system
yields the masses of both stars, as well as the three-dimensional
orbit and the distance to the system. This method has long been used
to measure the masses of stars in resolved binaries
\citep[e.g.][]{1969AJ.....74..689S} and extended to binaries resolved
by interferometry \citep[e.g.][]{2018AJ....156..185D}. A further
measurement of the radii of both stars would provide two mass-radius
points.

\cite{1971MNRAS.151..161H} already succeeded in one mass-radius
measurement. They observed the binary $\alpha\rm\,Vir$ (Spica) using
optical intensity interferometry at mas-scale resolution, as well as
spectroscopy. Combining these observations they inferred the distance,
the masses of both stars and the radius of the brighter star, as well
as all the orbital parameters (see their Table~III). Only the radius
of the second star was lacking. Unfortunately the progress of
intensity interferometry stopped soon after, as the interest of the
comunity shifted to other kinds of interferometry, and an
interferometric measurement of the radius of the second star still
remains to be done. However, a renewed interest in intensity
interferometry and new observations resolving single stars
\citep[e.g.,][]{2020MNRAS.491.1540A,2020MNRAS.494..218R,Abeysekara2020,2021arXiv210705596Z}
suggest that it will soon become possible to use this technique to do
mass-radius determinations in binary stars. The great advantage of
intensity interferometry is that it does not require sub-wavelength
precision on the mirrors or other optical components, and hence can be
extended to very long interferometric baselines. Baselines of
$>1\rm\,km$, corresponding to resolutions of $<1\rm\,mas$ are planned
\citep{DRAVINS2013331} and still longer baselines have been advocated
\citep{2020MNRAS.498.4577B}.

Intensity interferometry measures the squared visibility, or
equivalently the 2D angular power spectrum of the source brightness,
and from the observable it is possible to reconstruct the image of two
stars \citep{DRAVINS2013331}. In this work we adopt a different
strategy, which is to take a parameterised model for two limb-darkened
stars and fit their radii and separation to the interferometric
observable.

This work discusses parameter estimation using simulated data for a
binary.  A Spica-like system is taken as an example, but no special
properties of Spica are assumed.  We consider two methods to recover
the system parameters from the squared visibility.  The first involves
comparing the observable at two (or more) wavelengths to separate the
visibility contributions of the two stars.  Essential for this is that
the two stars have different effective temperatures, and hence
contribute differently at different wavelengths.  The second method
involves direct fitting of the squared visibility to a two-star model
with parameters for the stellar radii, the projected distance, and
limb darkening coefficient.

This paper is organized as follows. In Section~2, we describe the
methods of separation in visibilities and direct fitting to estimate
parameters, assuming a uniform-brightness model of the star. However,
in reality, stars do not follow a uniform model of brightness. So in
Section~3 we describe a limb darkening model of the stars. This is
followed in Section~4 by simulations of intensity-correlation data and
the recovery of the parameters of interest using a Markov chain Monte
Carlo (MCMC) technique.  The results our analysis are described in
Section 5. Finally we conclude in Section 6.

\section{Methodology}
An amplitude interferometer measures the modulus and phase of the Fourier transform of the brightness distribution of an incoherent source.  Hence the response of the amplitude interferometer, if sufficiently well sampled, can be inverted to obtain the original image of the source.  Techniques for doing so are well known, especially in radio-astronomy \citep[see e.g.,][section~10.4]{thompson2017interferometry}. The processing of an image obtained in this way, is heavily dependent on the ability of the measuring device to accurately track the phase of the light signal coming from sources.  The latter is not required in intensity interferometry, which measures only the modulus of the Fourier transform of a source’s brightness distribution. Because of the lack of phase information, reconstruction of a source having a complex structure requires information to be added, as a model or other prior. In the case where the source’s shape is known, a precise size estimate can be extracted in a model-dependent manner. 

\subsection{Correlated Signal}
We consider first a binary source comprising two stars $A$ and $B$ of uniform brightness with angular radii $\Theta_A$ and $\Theta_B$, and effective temperature $T_{A}$ and $T_{B}$ respectively, at angular distance $\Theta_d$ from each other.
The total flux of photons coming from the binary source is the weighted sum of the photon flux coming from the individual stars
\begin{eqnarray} \label{equ:totalflux}
  \Phi = I_A + I_B =
  |S_{A}(\Omega)|^2 \pi \Theta_A^2 + |S_{B}(\Omega)|^2 \pi \Theta_B^2
\end{eqnarray}
where $I_A=|S_{A}(\Omega)|^2$ and $I_B=|S_{B}(\Omega)|^2$ are the flux of photons from stars $A$ and $B$ respectively.

For a star of uniform brightness, visibility function at a point
$(u,v)$ on the interferometric plane has the well-known form
\begin{eqnarray}
  V_A(u,v) \propto \frac{J_1 (\rho\,\Theta_A)}{(\rho\,\Theta_A)}
\end{eqnarray}
and analogously for $V_B(u,v)$, where
\begin{eqnarray}
  \rho = \sqrt{u^2+v^2}
\end{eqnarray}
For both stars together the corresponding visibility function is
\begin{eqnarray}
    V(u,v) = \frac{\displaystyle{I_A V_A(u,v) + I_B V_B(u,v) \exp\left(i \rho\,\Theta_d\cos\psi \right)}}{I_A + I_B}
    \label{equ:Visibility_UV}
\end{eqnarray}
where the $u$ axis of the baseline is oriented along the line connecting the sources. The square of the modulus of the visibility function
\begin{equation}
    |V(u,v)|^2 = \frac{\Biggr[\splitfrac{I_A^2 V_A^2(u,v) + I_B^2 V_B^2(u,v)}{+2 I_A I_B V_A(u,v)V_B(u,v) \cos{\left( \rho\,\Theta_d\cos\psi \right)}}\Biggr]}{\left[I_A + I_B \right]^2}
    \label{equ:squared_visibility}
\end{equation}
is the main observable in intensity interferometry.
By construction $V(0,0)=1$.  In practice, however, the measured quantity is
\begin{eqnarray}
    g(u,v) = \frac{\Delta\tau}{\Delta t}|V(u,v)|^2
    \label{equ:HBT_corr}
\end{eqnarray}
where $\Delta\tau$ is the coherence time or equivalently the
reciprocal of the frequency bandwidth \citep{mandel1965coherence}
while $\Delta t$ is the time resolution.

\subsection{Methods to fit parameters}
The discussion of this paper is based on two possible strategies, which we now describe, to estimate the parameters of the stars.

\subsubsection{Separation in visibilities}

In this method, we consider measurements at two different wavelengths.  Provided the stars have different effective temperatures, the squared visibility will be measured with two different values of the flux ratio $f = I_B/I_A$. Then there is a possibility to estimate both radii of the system.

Along the $v$ axis, i.e., $u=0$ the visibility \eqref{equ:Visibility_UV} will be real.  We may then take the square root of the squared visibility \eqref{equ:squared_visibility} to obtain
\begin{eqnarray}
    V_1(0,v) = \frac{V_A(0,v) + f_1 V_B(0,v)}{1 + f_1}
    \label{equ:Visibility_1}
\end{eqnarray}
and
\begin{eqnarray}
    V_2(0,v) = \frac{V_A(0,v) + f_2 V_B(0,v)}{1 + f_2}
    \label{equ:Visibility_2}
\end{eqnarray}
Assuming the sign of $V_1(0,v)$ and $V_2(0,v)$ can be inferred, the visibility contributions of source $A$ and source $B$ in terms of $V_1(0,v)$ and $V_2(0,v)$ are
\begin{eqnarray}
    V_A(0,v) = \frac{\Biggr[\splitfrac{V_1(0,v) f_1 f_2 - V_2(0,v) f_1f_2} {+ V_1(0,v)f_2 - V_2(0,v)f_1}\Biggr]}{f_2 - f_1}
    \label{equ:Visibility_SEP_A}
\end{eqnarray}
and
\begin{eqnarray}
    V_B(0,v) = \frac{\Biggr[\splitfrac{f_1 V_1(0,v) - f_2 V_2(0,v)}{ + V_1(0,v) - V_2(0,v)}\Biggr]}{f_1 - f_2}
    \label{equ:Visibility_SEP_B}
\end{eqnarray}
The expressions \eqref{equ:Visibility_SEP_A} and \eqref{equ:Visibility_SEP_B} can now be used to fit parameters, specifically the radii of both stars. But using this method, the distance between stars can not estimated as the above equation are independent of $\Theta_d$.

\subsubsection{Direct fitting}
The second method we consider is directly fitting the 
squared visibility \eqref{equ:squared_visibility} to infer the parameters.
Since $|V(u,v)|^2$ depends explicitly on $\Theta_d$ as well as on $\Theta_A$ and $\Theta_B$, we can use this correlation to estimate the angular separation between the sources together with angular radii of the sources.

\subsection{Slope of the Signal}

In order to better understand the dependence of the squared visibility on each of the radii in different parts of the $u,v$ plane, we consider derivatives of the former (equ.~\ref{equ:squared_visibility}) with respect to $\Theta_A$ and $\Theta_B$ (taking $I_A$ and $I_B$ as normalised constants) for $u=0$, which are as follows.
\begin{eqnarray}
    \frac{\partial|V(0,v)|^2}{\partial\Theta_A} = -\frac{4I_{A}V(0,v)}{(I_A + I_B)}\frac{J_2\left(v\Theta_A \right)}{\Theta_A}
    \label{equ:change_visi_A}
\end{eqnarray}
and 
\begin{eqnarray}
    \frac{\partial|V(0,v)|^2}{\partial\Theta_B} = -\frac{4I_{B}V(0,v)}{(I_A + I_B)}\frac{J_2\left(v\Theta_B \right)}{\Theta_B}
    \label{equ:change_visi_B}
\end{eqnarray}
The first factor of eqns.~\eqref{equ:change_visi_A} and \eqref{equ:change_visi_B} depends on both sources. However, the second factor depends on only the source with respect to which equ.~\eqref{equ:squared_visibility} is differentiated.  As we can see, the derivative is a product of Bessel functions $J_1$ and $J_2$, and can be positive or negative. So, there will be some regions in the $u,v$ plane where the derivative of the observable will be positive and some regions where it will be negative. Using these equations, we can remove the data from the baseline region that reduces the strength of the signal (i.e., the area of baselines from where signals with negative slope observed).

\section{Impact of Limb-Darkening}
Limb darkening is the phenomenon that describes the brightness of a star as a function of its radius, observed relative to the maximum brightness at the center of the stellar disk. The disk of the star is dimmer towards the limbs than at the center. As first explained by \cite{schwarzschild1906gleichgewicht} limb darkening occurs due to the gradient of temperatures through the layers of the star.  Higher temperatures emit more photons at a given wavelength, and lower temperatures emit fewer. The photons from a star are emitted from one optical depth within the surface of the star. Because of the star's geometry, one optical depth at the center of the stellar disk penetrates through to a deeper layer of the star than one optical depth near the limb; therefore, we see a hotter region, and thus middle of the stellar disk appears brighter.

Limb darkening is typically treated as a simple parameterised function of the angle between a line normal to the stellar surface and the observer's line of sight, which makes fitting the stellar intensity profile much simpler and reduces the number of free parameters. The most common parametrizations are linear and quadratic relations \citep{1950HarCi.454....1K}, but other suggested relations include a four-parameter relation, a square-root relation as well as exponential and logarithmic relations \citep{10.1093/mnras/81.5.361,1992A&A...259..227D,1970AJ.....75..175K,2003A&A...412..241C,2009A&A...505..891S,2000A&A...363.1081C}. Following the work described in \cite{10.1093/mnras/167.3.475} we have used  a limb darkening model of the form
\begin{eqnarray}
   \left|\frac{S(\Omega,\mu)}{S(\Omega,1)}\right|^{2} = 1-l(1-\mu)-m(1-\mu)^2-n(1-\mu)^3
   \label{equ:ld_nonlinear}
\end{eqnarray}
where $l$, $m$ and $n$ are limb coefficient depending on the surface temperature and frequency, and $\mu$ is the cosine of the angle $\theta$ between the normal to the surface at that point and the line of sight from the star to the observer. In terms of the radius $R$ of the star, the projected radial position $R^{\prime}$ of the observed point, and the angle $\theta$ is expressed as
\begin{eqnarray}
    \mu = \cos\theta = \sqrt{1- \frac{R'^2}{R^2}}
    \label{equ:cosine}
\end{eqnarray}
Given the limb darkening model under consideration and the under the assumptions that the values of the limb-darkening coefficients are frequency independent and same for both the stars, the total flux of binary source is given by
\begin{eqnarray}
    \Phi = C \left(I_A + I_B\right) 
    \label{equ:ld_totalflux}
\end{eqnarray}
where the constant $C$ in terms of limb coefficients is given by,
\begin{eqnarray}
    C = \left(1-\frac{l}{3}-\frac{m}{6}-\frac{n}{10}\right)
    \label{equ:extrafactor}
\end{eqnarray}
and the visibility functions of star $A$ and $B$ are given by
\begin{eqnarray}
    V_A(u,v) = \sum_i a_i \frac{J_i (\rho\,\Theta_A)}{(\rho\,\Theta_A)^i}
    \label{equ:visibility_A_ld}
\end{eqnarray}
\begin{eqnarray}
    V_B(u,v) = \sum_i a_i \frac{J_i (\rho\,\Theta_B)}{(\rho\,\Theta_B)^i}
    \label{equ:visibility_B_ld}
\end{eqnarray}
For $i=1,~3/2,~2,~5/2$ and the value of $a_i$ for each $i$ are
\begin{eqnarray}
\begin{aligned}
     &a_1 = 1-l-m-n \\
     &a_{3/2} = \sqrt{\frac{\pi}{2}}(l+2m+3n) \\
     &a_2 = -2(m+3n) \\
     &a_{5/2} = 3 \sqrt{\frac{\pi}{2}}n
     \label{equ:condition_ld}
\end{aligned}
\end{eqnarray}
In the absence of star $B$, the result will be according to \cite{10.1093/mnras/167.3.475}.

Eqns.~\eqref{equ:visibility_A_ld} and \eqref{equ:visibility_B_ld} are the generalized form of visibilities of sources $A$ and $B$, according to the coefficient parameters $l, m, n$ the limb darkening model of star assumed here.

\section{Simulations and parameter fitting}

We now proceed to simulate intensity-correlation data and recover the parameters using a Markov chain Monte Carlo (MCMC) algorithm.

\subsection{Signal to noise}

We may write the signal as
\begin{eqnarray}
     g(u,v,\omega) = \frac{\Delta\tau}{\Delta t} \, |V(u,v,\omega)|^2
     \label{equ:HBT_corr_MCMC}
\end{eqnarray}
which is equ.~\eqref{equ:HBT_corr} slightly rewritten with $\omega$
standing for all the parameters of the system (radii, distance, and
limb-darkening coefficients).  For the signal at the $i$-th baseline,
we write
\begin{eqnarray}
    g_i(u_i,v_i, \omega) = \frac{\Delta\tau}{\Delta t} \, |V(u_i,v_i,\omega)|^2
    \label{equ:HBT_corr_i}
\end{eqnarray}
and the data collected at that baseline will be
\begin{eqnarray}
  d_i = g_i + \sigma_i {\cal N}_i 
  \label{equ:data_i}
\end{eqnarray}
where $\sigma_i$ is the noise level and ${\cal N}_i$ is a Gaussian
random number.\footnote{The noise level depends on the observing time at
  each baseline, and in the simulations we assume all the $\sigma_i$
  to the same for all baselines, though this is not essential.}  The
signal-to-noise at the $i$-th baseline will be
\begin{eqnarray}
  {\rm SNR}_i = \frac{d_i}{\sigma_i}
  \label{equ:SNR_i}
\end{eqnarray}
while
\begin{eqnarray}
  {\rm SNR} = \Big( {\textstyle\sum_i} \, {\rm SNR}_i^2 \Big)^{1/2}
  \label{equ:SNR_total}
\end{eqnarray}
will be the total signal-to-noise ratio for a full simulation.

For the visibility-separation method, the relation between ${\rm
  SNR}_i$ and $\sigma_i$ is more complicated.  First, the observable squared visibility has to be converted to individual
visibilities using eqns.~\eqref{equ:Visibility_1} and
\eqref{equ:Visibility_2}.  Correspondingly, the ${\rm SNR}_{|V|^2}$
(say) relating to the observable, has to be converted to an effective
\begin{eqnarray}\label{equ:snrvv2}
    {\rm SNR}_V = 2{\rm SNR}_{|V|^2}
\end{eqnarray}
using the error propagation formula.  Further using the latter formula
we have
\begin{eqnarray}
\sigma_{V_A}^2 =
\left(\frac{\partial V_A}{\partial V_1}\right)^2 \sigma_{V_1}^2 +
\left(\frac{\partial V_A}{\partial V_2}\right)^2 \sigma_{V_2}^2 
\end{eqnarray}
and similarly $\sigma_{V_B}^2$.  The values of $\sigma_{V_A,i}$ and
$\sigma_{V_B,i}$ are required for parameter-fitting.

To relate the SNR in a simulation to the expected SNR in an observing
run, we recall that for one baseline
\begin{eqnarray}\label{equ:snr}
  {\rm SNR}_i \approx \left(A_1 A_2\right)^{1/2} \,
  \Phi \, |V(u_i,v_i)|^2 \left(t^{\rm obs}_i/\Delta t\right)^{1/2}
\end{eqnarray}
where $A_1$ and $A_2$ are the effective collecting areas,
\begin{eqnarray}\label{equ:specAB}
  \Phi \approx \frac\lambda{1\mu\rm m} \times 10^{-4-{\rm AB}/2.5} 
       \; \hbox{photons} \; {\rm m^{-2} \, s^{-1} \, Hz^{-1}}
\end{eqnarray}
is the spectral photon density per polarisation
  channel, and $t^{\rm obs}_i$ is the observing time at that
baseline.  The full SNR of the observing run is then obtained using
equ.~\eqref{equ:SNR_total} as before.

A counterintuitive fact here is that while the
  observable $g$ depends on the coherence time $\Delta\tau$, the SNR
  is independent of $\Delta\tau$.  This was first noted by
  \cite{1957RSPSA.242..300B} and can be understood as follows.
  Suppose we change the bandwidth by a factor of $k$, while keeping
  the detector time-resolution $\Delta t$ fixed.  This is equivalent
  to changing the $\Delta\tau$ by $1/k$.  There will be $k$ times as many
  photons in each detector per $\Delta t$, and hence $k$ times as many
  intensity-interferometric correlations --- not $k^2$ times, because
  photons in different slices of the band do not interfere.
  However, the $k$ times as many photons will have $k^2$ times as many chance
  coincidences, implying $k$ times as much shot noise.  As a result
  both signal and noise will change by the same factor $k$.
We remark that all of this is valid only if the
  coherence time $\Delta\tau$ is much smaller than the instrumental
  time resolution $\Delta t$, which is a safe assumption for any
  foreseeable instrumentation. Furthermore, for intensity
  interferometry with three-photon or higher-order correlations the
  SNR does depend on $\Delta\tau$ \cite[see
    e.g.,][]{2014MNRAS.437..798M}.

\subsection{Parameter fitting}

For a given data set the likelihood in the usual Bayesian formulation
will be
\begin{eqnarray}
    \ln P \left ( \{d_i\}|\alpha,\omega \right ) =
-{\textstyle\frac12} \sum_i \sigma_i^{-2}
\Big( d_i - \alpha|V(u_i,v_i,\omega)|^2 \Big)^2
    \label{equ:lnP_corr}
\end{eqnarray}
where $\alpha=\Delta\tau/\Delta t$ is an additional parameter, which
will not be accurately known in advance.  We can marginalise out the
nuisance parameter $\alpha$ to avoid having to fit it, as follows.
Assuming a flat prior for $\alpha$ we have
\begin{eqnarray}
    P \left (\{d_i\}|\omega\right) = \int P \left (\{d_i\}|\alpha,\omega\right) d\alpha
    \label{equ:P_corr}
\end{eqnarray}
We now define two scalar products: first of the data with the model
\begin{eqnarray}
    G = \sum_i \sigma_i^{-2} \, d_i \, |V(u_i,v_i,\omega)|^2
    \label{equ:likli_G}
\end{eqnarray}
and, the scalar product of model with itself
\begin{eqnarray}
    W = \sum_i \sigma_i^{-2} \, |V(u_i,v_i,\omega)|^4
    \label{equ:likli_W}
\end{eqnarray}
The integral \eqref{equ:lnP_corr} can now be expressed as
\begin{eqnarray}
    \ln P\left( \{d_i\}|\omega\right) = \frac{G^2}{2 W} - \frac12\ln W 
    \label{equ:ln_likli}
\end{eqnarray}

Eqns.~(\ref{equ:likli_G}--\ref{equ:ln_likli}) are convenient for input
to MCMC.  An MCMC implementation, such as the well-known emcee code
\citep{2013PASP..125..306F} which we used, generates a sample of
points $\omega$ in parameter space sampled according to some given
function of $\omega$.  Sampling is desired according to
$P\left(\{d_i\}|\omega\right)$ times the prior probability of
$\omega$, the latter representing the allowed ranges of the various
parameters.  The resulting sample of $\omega$ points cannot in general
be visualised directly, and hence projections to one- and
two-parameter subspaces are conventionally used, as in
Fig.~\ref{fig:corner_method2}.

\begin{figure*}
    \centering
\begin{subfigure}{0.50\textwidth}
    \includegraphics[width=\linewidth]{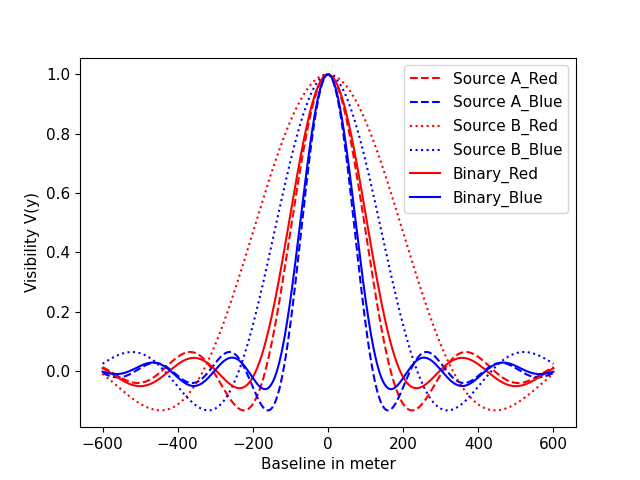}
    \caption{Visibility along the baseline}
    \label{fig:visibility_meter}
\end{subfigure}\hfil 
\begin{subfigure}{0.50\textwidth}
    \includegraphics[width=\linewidth]{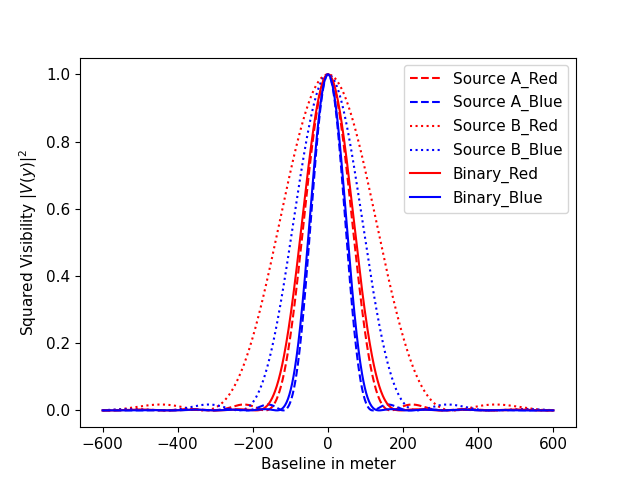}
    \caption{Squared visibility along baseline}
    \label{fig:squa_visibility_meter}
\end{subfigure}\hfil 
\begin{subfigure}{0.50\textwidth}
    \includegraphics[width=\linewidth]{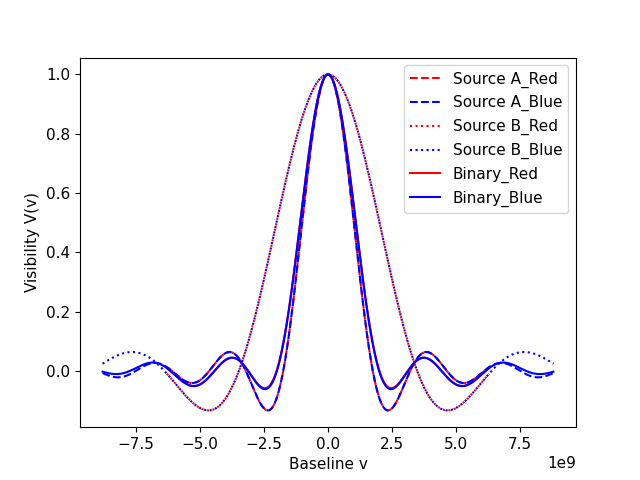}
    \caption{Visibility along unit-less baseline}
    \label{fig:visibility_unit_less}
\end{subfigure}\hfil 
\begin{subfigure}{0.50\textwidth}
    \includegraphics[width=\linewidth]{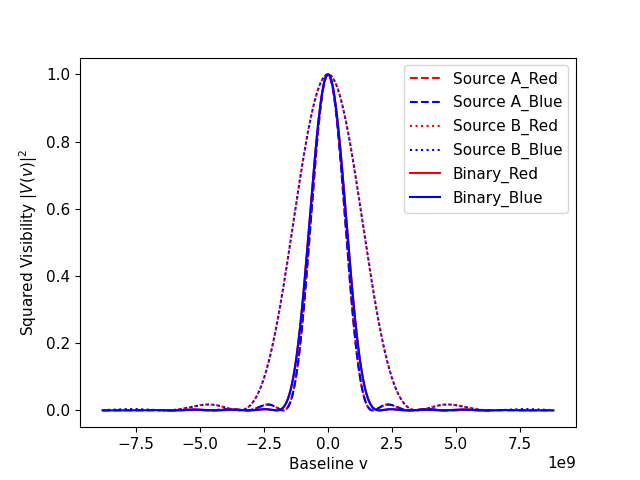}
    \caption{Squared visibility along unit-less baseline}
    \label{fig:squa_visibility_unitless}
\end{subfigure}\hfil 
\caption{Analytical plot of the signal at two different wavelengths, for baselines perpendicular to the line between the stars. The two left panels show the visibility, while the two right panels show the squared visibility.  For the two upper panels the baseline is in metres, while the two lower panels use the dimensionless baseline $v$.  According to eqns.~\eqref{equ:Visibility_UV} and \eqref{equ:squared_visibility} the signal is always unity at zero baseline.}
    \label{fig:Signal}
\end{figure*}

\begin{figure}
\subfloat[SNR = 32700]{
  \includegraphics[width=\columnwidth]{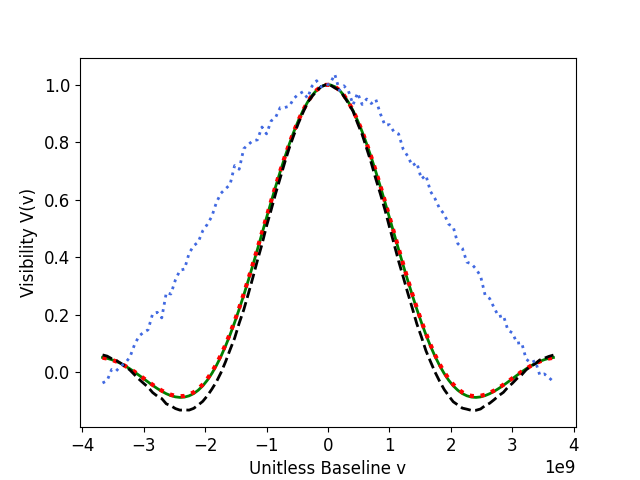}
}
  \label{fig:visi_1e-4}
\subfloat[SNR =650]{
  \includegraphics[width=\columnwidth]{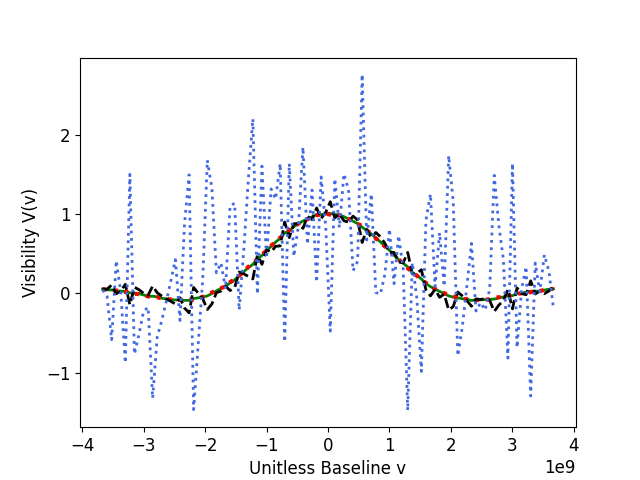}%
}
  \label{fig:visi_5e-3}
\subfloat[SNR =320]{
  \includegraphics[width=\columnwidth]{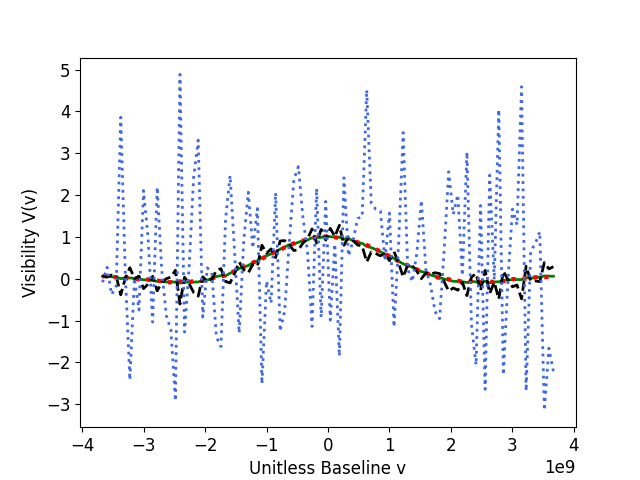}
}
  \label{fig:visi_1e-2}
\caption{Visibility separation from simulated data illustrated at different SNR levels.  The dotted red and plane green line show the
  combined visibility with $f_1=0.1$ and $f_2=0.11$ respectively (see
  eqns.~\ref{equ:Visibility_1} and \ref{equ:Visibility_2}).  The
  baseline $v$ is perpendicular to the line between the two sources.
  The dashed black line and the dotted royalblue line show the inferred
  individual visibilities of source $A$ and source $B$
  respectively.}
  \label{fig:visi_sep}
\end{figure}

\begin{figure}
\subfloat[Corner plot for SNR=32700]{
  \includegraphics[width=\columnwidth,height=7.2cm]{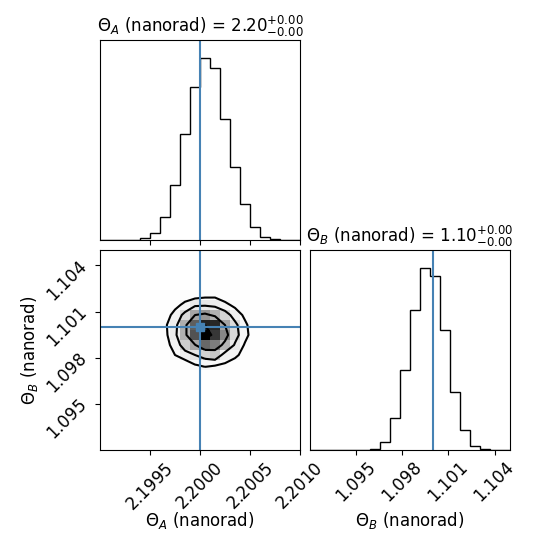}
}
  \label{fig:corner_32700_method1}

\subfloat[Corner plot for SNR=650]{
  \includegraphics[,width=\columnwidth,height=7.2cm]{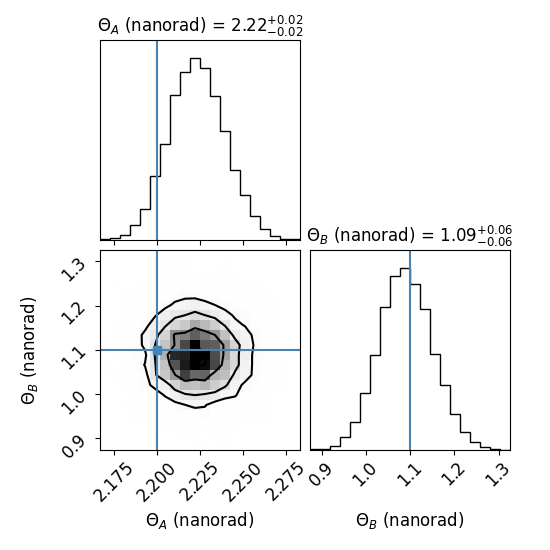}}
  \label{fig:corner_650_method1}
  
\subfloat[Corner plot for SNR=320]{
  \includegraphics[width=\columnwidth,height=7.2cm]{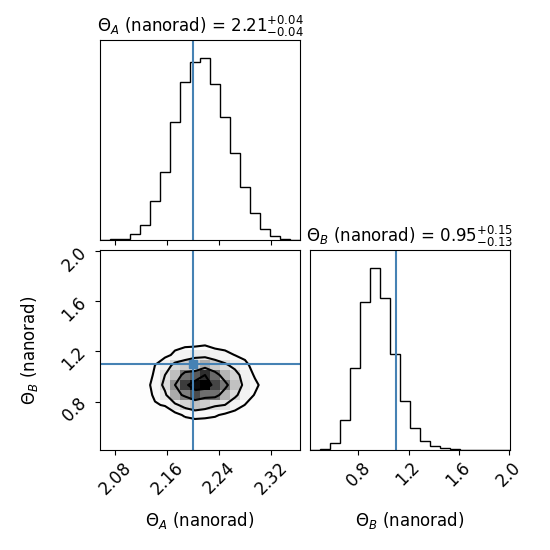}}
  \label{fig:corner_320_method1}
\caption{Estimation of the angular radii (in nanorad) of stars $A$ and $B$ using separated visibilities shown in Fig.~\ref{fig:visi_sep}. \label{fig:corner_method1}}
\end{figure}

\begin{figure*}
    \centering
\begin{subfigure}{0.50\textwidth}
    \includegraphics[width=\linewidth]{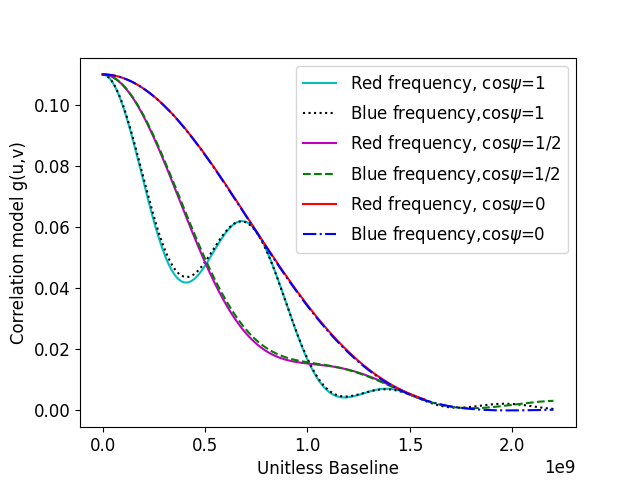}
    \caption{Model data along baseline}
    \label{fig:HBT_model}
\end{subfigure}\hfil 
\begin{subfigure}{0.50\textwidth}
    \includegraphics[width=\linewidth]{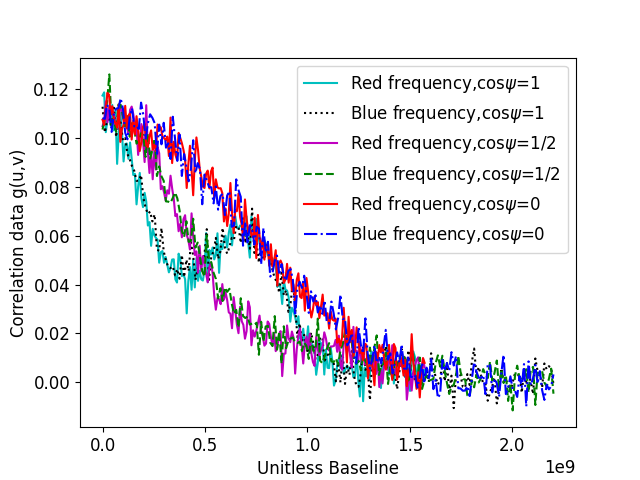}
    \caption{Simulated data along baseline}
    \label{fig:HBT_data}
\end{subfigure}\hfil 
\caption{Observable intensity correlation $g(u,v)$ at
  differently-oriented baselines and different observing wavelengths.
  The labelled $\cos\psi=u/\sqrt{u^2+v^2}$ where $u$ is along the line
  connecting the two stars.  The left panel shows the model according
  to eqns.~\eqref{equ:squared_visibility} and \eqref{equ:HBT_corr}.
  The right panel shows the signal with added noise corresponding to a
  total SNR of 370. \label{fig:HBT_corr}}
\end{figure*}
\begin{figure}
    \centering
    \includegraphics[width=\linewidth]{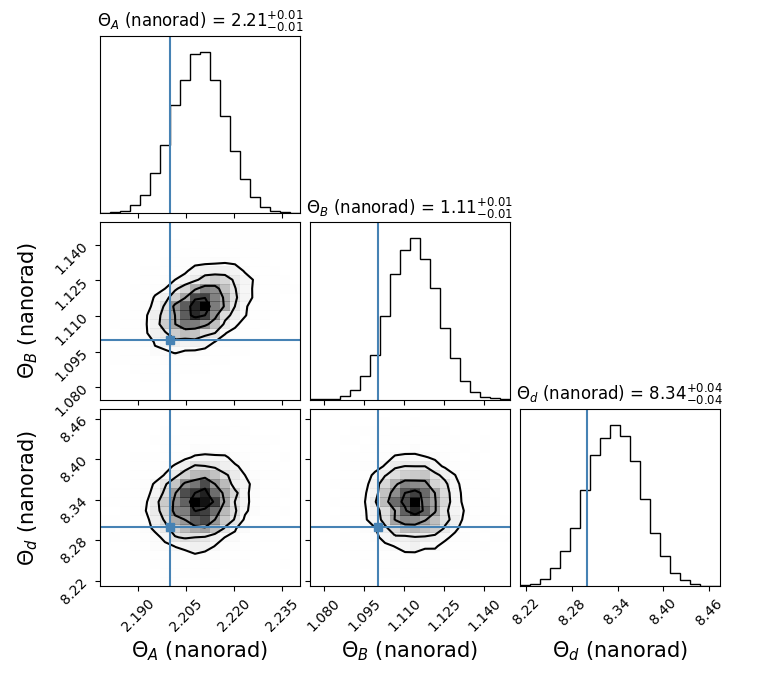}
    \caption{Estimation of the angular radii of the two stars and
      angular separation between them using the direct fitting method,
      from the simulated data shown in Fig.~\ref{fig:HBT_data}. Note
      the positive correlation between angular radii of the two sources.}
    \label{fig:corner_method2}
\end{figure}
\begin{figure}
    \centering
    \includegraphics[width=\linewidth,]{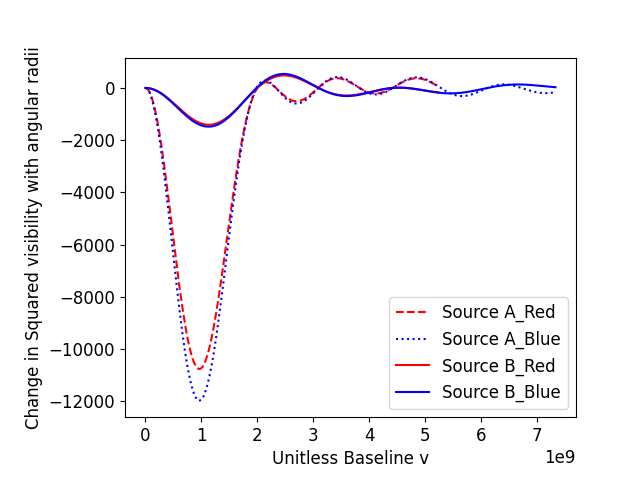}
    \caption{Derivatives of the squared visibility with respect to
      angular radii of the stars (eqns.~\eqref{equ:change_visi_A} and
      \eqref{equ:change_visi_B}), shown along baselines perpendicular to
      the line connecting the stars.  The regions with a sharp negative slope
      in the signal are considered for exclusion in the following
      simulations.}
    \label{fig:slope_visi}
\end{figure} 
\begin{figure*}
    \centering
\begin{subfigure}{0.50\textwidth}
    \includegraphics[width=\linewidth]{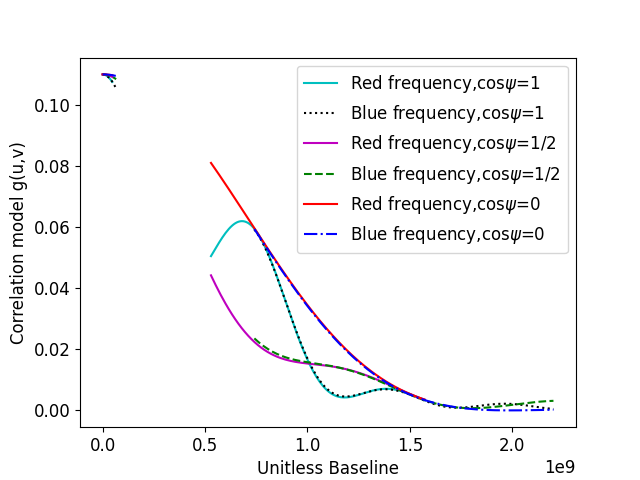}
    \caption{Model data along baseline}
    \label{fig:HBT_model_msk}
\end{subfigure}\hfil 
\begin{subfigure}{0.50\textwidth}
    \includegraphics[width=\linewidth]{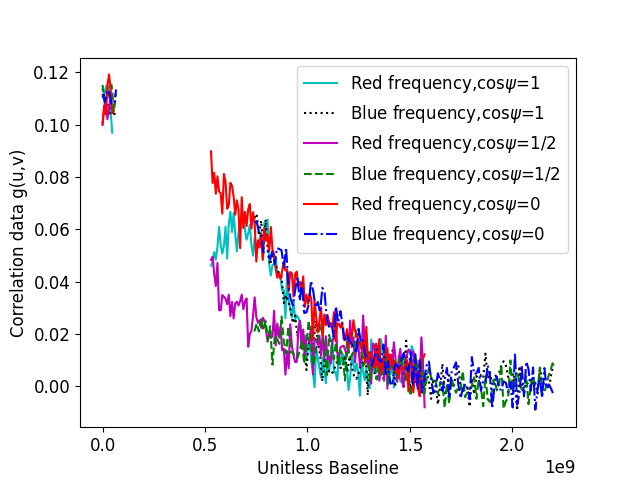}
    \caption{Simulation data along baseline}
    \label{fig:HBT_data_msk}
\end{subfigure}\hfil 
    \caption{Observable HBT correlation $g(u,v)$ after removing the
      region of baseline identified in Fig.~\ref{fig:slope_visi}.  As
      Fig.~\ref{fig:HBT_corr} the left panel is for the model signal
      and the right panel adds noise.  The total SNR here is 208.}
    \label{fig:HBT_corr_msk}
\end{figure*}

\section{Results}

In this section we first describe the signal of our interest i.e., visibility and squared visibility. This is followed by the estimation of parameters in two ways: first by separation of visibilities, and second the direct fitting of the squared visibility.

The parameter values used for the simulations are $\Theta_A = 2.20,
\Theta_B = 1.10$ nanoradians for the angular radii and $\Theta_d =
8.30$ nanoradians for the angular separation.\footnote{Recall that a
  mas is about 5~nanoradians.}  These values correspond roughly to the
Spica system \cite{1971MNRAS.151..161H,2016MNRAS.458.1964T}.
Simulated observations at $\nu=\SI{5e14}{\hertz}$ and
$\nu=\SI{7e14}{\hertz}$ are considered.  For brevity we refer to these
as red and blue.

The ratio $\Delta \tau/\Delta t$ of coherence time to resolution time
is taken to be $0.11$.  The latter may be much
  lower in practice.  For example, \citep{2021arXiv210705596Z}
  observed with $(\lambda,\Delta\lambda)$ values of
  $\rm(656.7\,nm,3\,nm)$ and $\rm(510.5\,nm,0.3\,nm)$ amounting to
  $\Delta\tau=\lambda^2/(c\Delta\lambda)$ of about $0.5\rm\,ps$ and
  $3\rm\,ps$ respectively, whereas $\Delta t\approx400\rm\,ps$.
  However, as explained in the previous section, the SNR does not
  depend on $\Delta \tau$, and in the simulations $\Delta\tau/\Delta
  t$ is simply an arbitrary normalisation which gets marginalised out
  as a nuisance parameter.

\subsection{Signal}

Fig.~\ref{fig:Signal} shows the main signal for baselines perpendicular to the orientation of the binary. As expected, the signal attains its maximum value for zero baseline, and for a single star the larger the source size the narrower is the visibility pattern. However, the visibility of the total system is weighted towards the brighter source. Source~A dominates the brightness both in blue and red, but is slightly less dominant in red; this difference is, however, not discernible from the figure. For certain wavelength (most noticibly around $175$m at blue wavelength) the visibility $V(0,v)$ is found to be less for the binary system than those for either of the sources under consideration. This is due to the fact that the visibilities for sources have  opposite sign and are interfering destructively.

\subsection{Measuring Parameters}
For visibility separation, the values of the parameters $f_1$ and $f_{2}$ were assumed to be $0.10$ and $0.11$ respectively. The separation has been done for different values of SNR as shown in Fig.~\ref{fig:visi_sep}.  The resulting estimates of the angular radii are shown in Fig.~\ref{fig:corner_method1}. Our results show that the minimum value of SNR for which this method works is around $300$.  

\begin{figure}
    \centering
    \includegraphics[width=\linewidth]{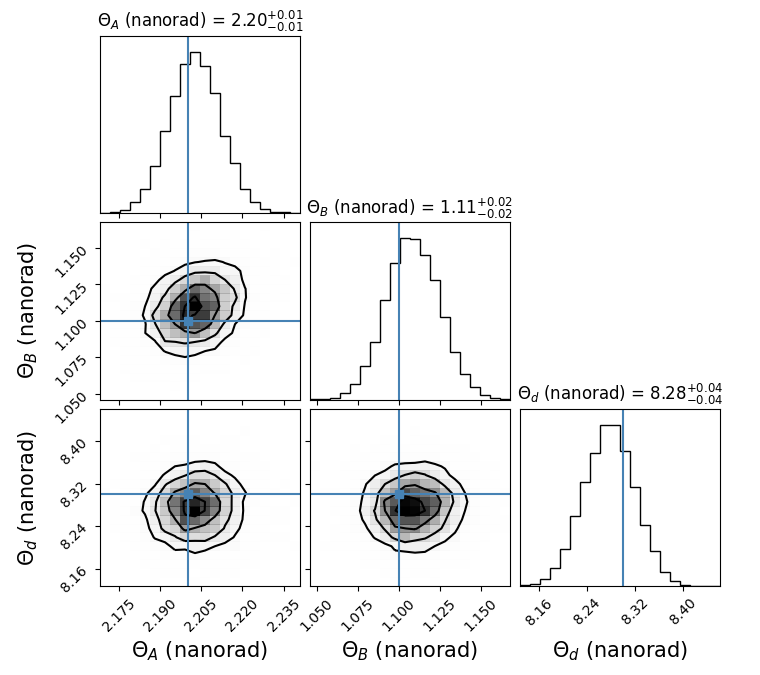}
    \caption{Estimation of the angular radii of the two stars and
      angular separation between them using the direct fitting method,
      from the simulated data shown in Fig.~\ref{fig:HBT_data_msk}.
      The difference from Fig.~\ref{fig:corner_method2} is that some
      baselines regions have been excluded, with the result that the
      positive correlation between inferred angular radii of both
      sources has been mostly removed. \label{fig:corner_method2_msk}}
\end{figure}

For direct fitting of the squared visibility, we simulated the HBT correlation (Fig.~\ref{fig:HBT_model}) along the baseline for $\cos\psi = 0, 0.5, 1$.  It is not necessary to have the signal along straight lines in the $(u,v)$ plane, but this choice conveniently illustrates the oscillatory features in the correlation, which are due to the presence of a cosine factor in the expression of correlation equ.~\eqref{equ:squared_visibility}. The added noise is such that the total ${\rm SNR}=370$. Fig.~\ref{fig:corner_method2} clearly shows good agreement between input and recovered values of the parameters.
Although Fig.~\ref{fig:corner_method2} shows no correlation between
angular separation and angular radii of either star and, however, it
shows positive correlation between the angular radii of the stars. In
order to explain the origin of the correlation between the angular
radii of the stars, we have plotted in Fig.~\ref{fig:slope_visi} the
change in squared visibility with angular radii of sources along
baseline for the orientation of the sources in the direction
perpendicular to the baseline. The sharp negative slope in
Fig.~\ref{fig:slope_visi} appears to be the cause of the correlation
between estimated angular radii of the stars. In order to resolve this
issue we tried removing from our analysis the unwanted portion of the
baseline and followed the same methodology to estimate the parameters
from the unmasked portion of the baseline, as shown in
Fig.~\ref{fig:HBT_corr_msk}.  The results in
Fig.~\ref{fig:corner_method2_msk} show no correlation between the
angular radii of the stars.
\begin{figure*}
    \centering 
\begin{subfigure}{0.50\textwidth}
    \includegraphics[width=\linewidth,height=7.2cm]{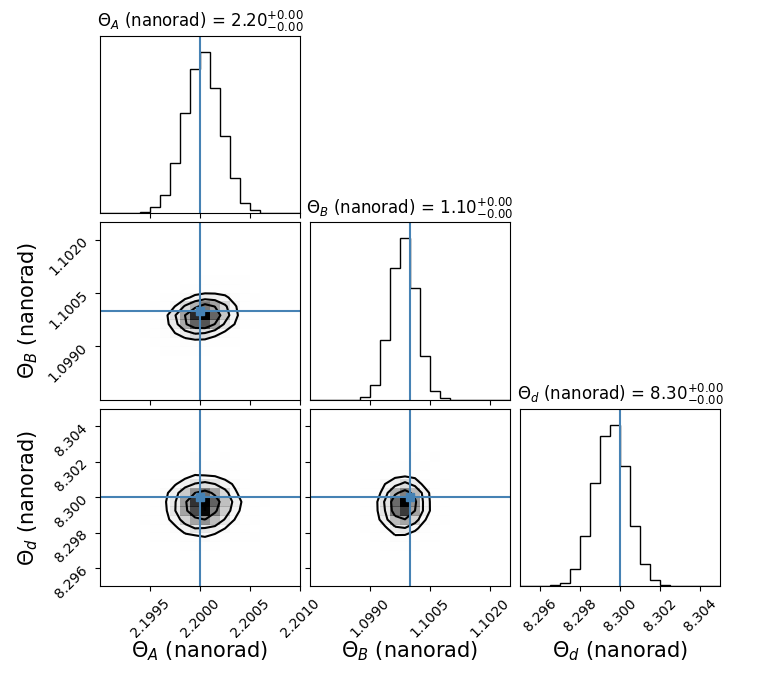}
    \caption{SNR = 10346}
    \label{fig:thresh_10346.54}
\end{subfigure}\hfil 
\begin{subfigure}{0.50\textwidth}
  \includegraphics[width=\linewidth,height=7.2cm]{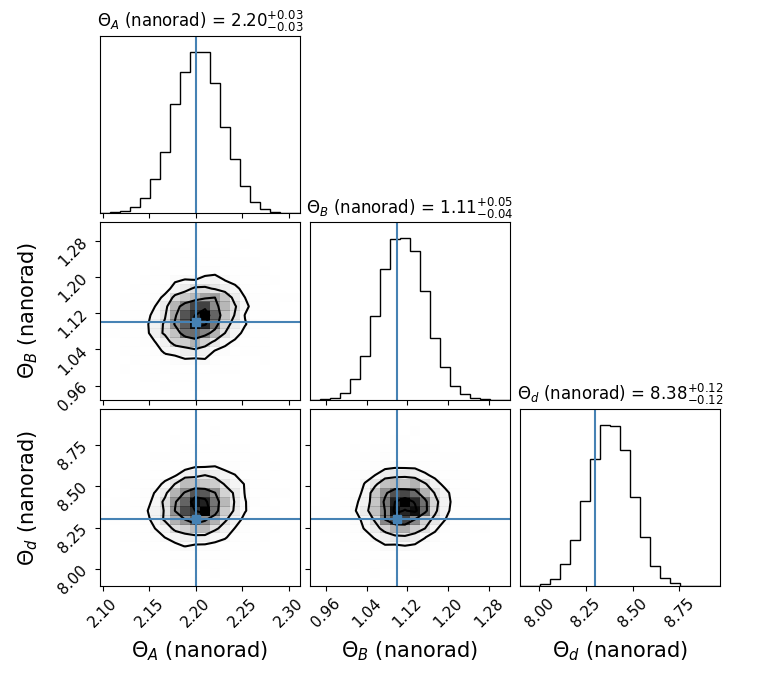}
  \caption{SNR = 79}
  \label{fig:thresh_79.27}
\end{subfigure}
\begin{subfigure}{0.50\textwidth}
    \includegraphics[width=\linewidth,height=7.2cm]{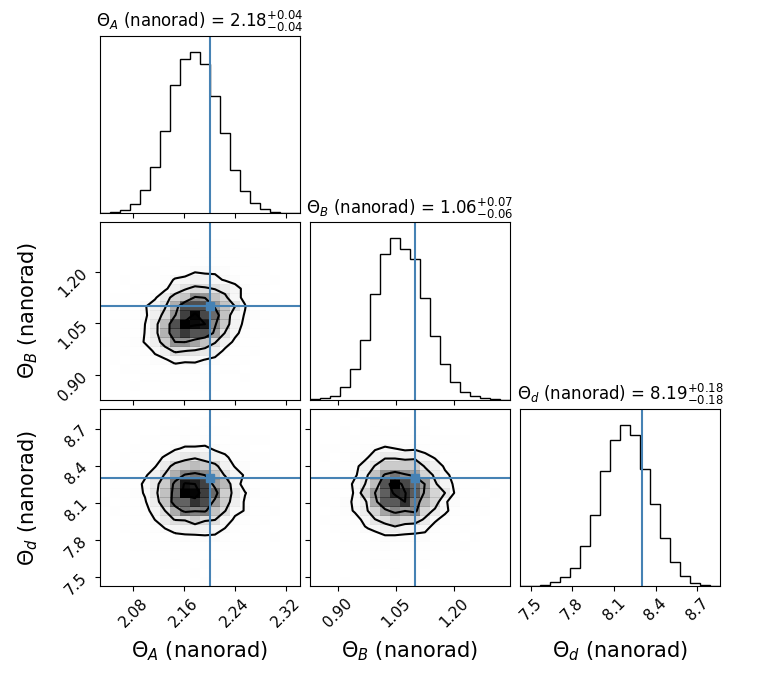}
    \caption{SNR = 55}
    \label{fig:thresh_55}
\end{subfigure}\hfil 
\begin{subfigure}{0.50\textwidth}
  \includegraphics[width=\linewidth,height=7.2cm]{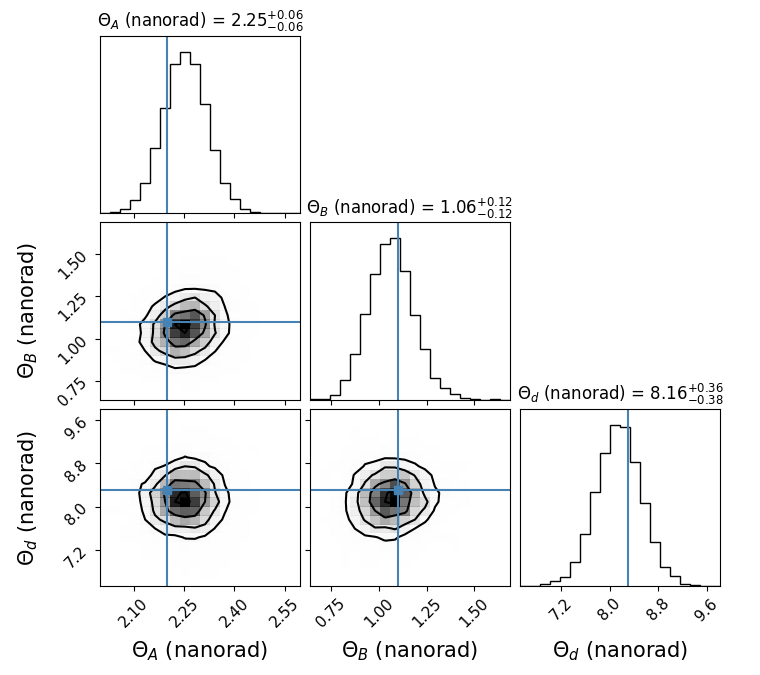}
  \caption{SNR = 41}
  \label{fig:thresh_41.23}
\end{subfigure}\hfil 
\begin{subfigure}{0.50\textwidth}
  \includegraphics[width=\linewidth,height=7.2cm]{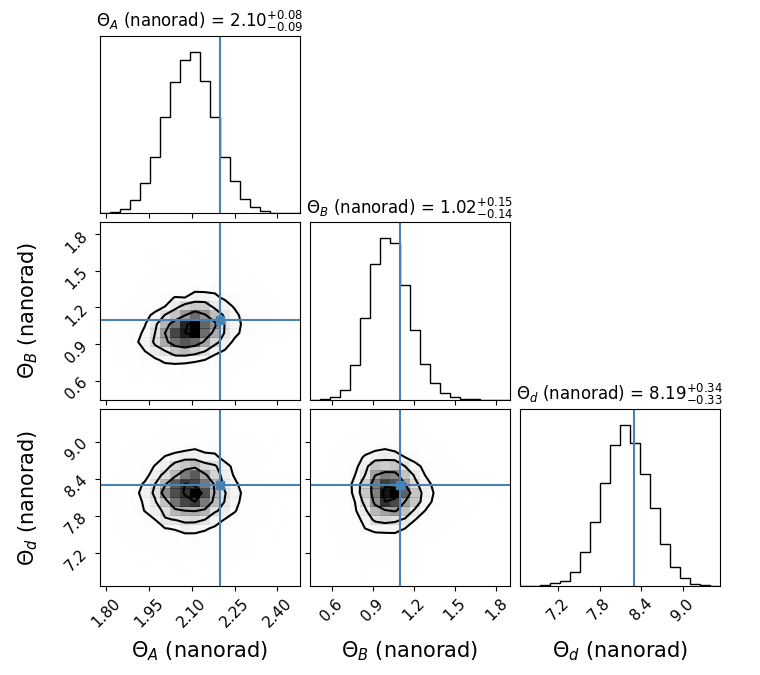}
  \caption{SNR = 35}
  \label{fig:thresh_35.51}
\end{subfigure}\hfil 
\begin{subfigure}{0.50\textwidth}
  \includegraphics[width=\linewidth,height=7.2cm]{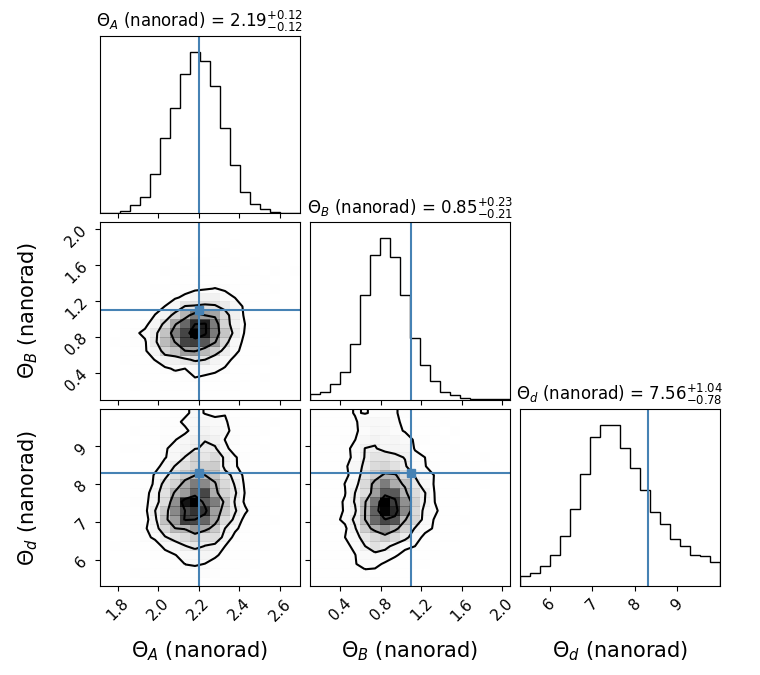}
  \caption{SNR = 32}
  \label{fig:thresh_32.30}
\end{subfigure}
  \caption{Estimations of the angular radii of both sources and the
    angular separation between them, using direct fitting at different
    total SNR. The threshold for estimating all three parameters is
    total SNR of about 30.}
  \label{fig:thresh_signal_to_noise}
\end{figure*}

\subsection{Threshold Signal-To-Noise Ratio}
Estimation of parameters works requires a minimum value of
the signal-to-noise ratio.  Fig.~\ref{fig:thresh_signal_to_noise}
checks this threshold value of the SNR for estimation of angular radii
of both sources and angular separation between them using a direct
fitting with masking unwanted region. The minimum value of SNR is
around 30.  If the total SNR is much lower than in
Fig.~\ref{fig:thresh_32.30}, parameter estimation becomes poor, with
the uncertainties becoming large and the credible regions excluding
the true value.

\begin{figure}
\subfloat[limb coefficients $l=0.6, m=0.0, n=0.0$]{
    \includegraphics[width=\columnwidth]{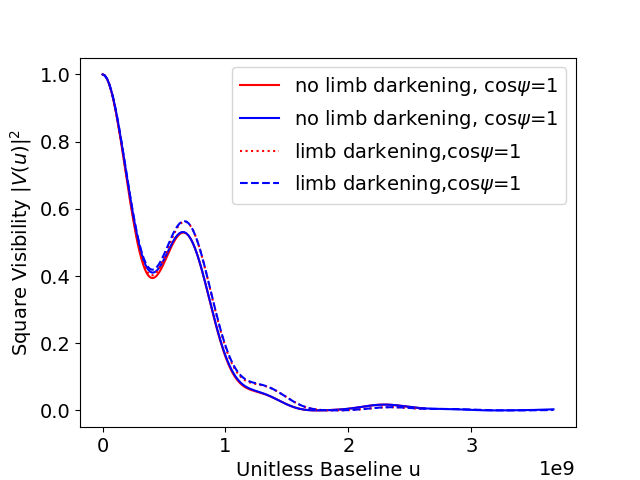}}
    \label{fig:effect_ld_linear}
\subfloat[limb coefficients $l=0.6, m=0.2, n=0.0$]{
    \includegraphics[width=\columnwidth]{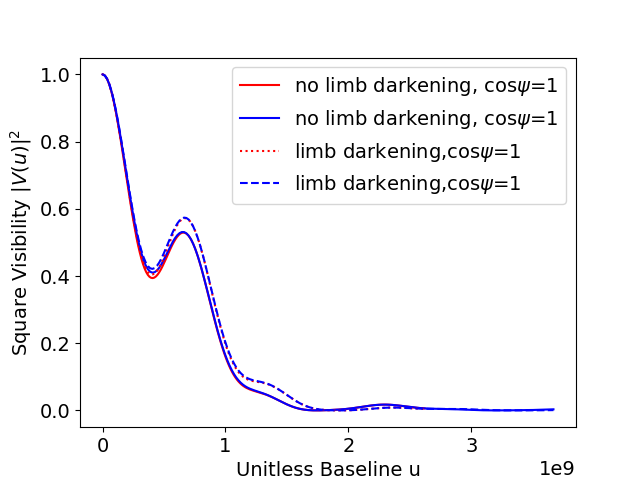}}
    \label{fig:effect_ld_quadretic}
\subfloat[limb coefficients $l=0.6, m=0.2, n=0.1$]{
    \includegraphics[width=\columnwidth]{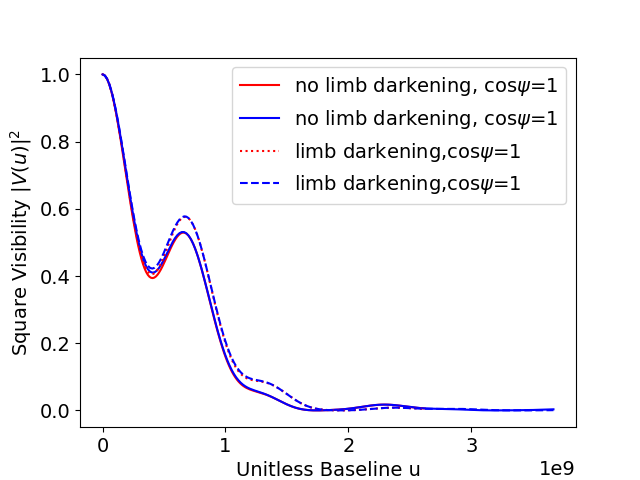}}
    \label{fig:effect_ld_nonlinear}
\caption{Analytical plots of the squared visibility function,
      comparing a uniform-brightness model and limb-darkened
      models. The effect of limb darkening is measurable only near the
      peaks of the signal curve. The three panels show the differences
      between linear, quadratic and cubic limb-darkening, which are
      very small. \label{fig:effect_ld}}
\end{figure}

\subsection{Effect of limb-darkening}
In order to show the impact of limb-darkening on the squared
visibility function of the stars visibilities for the orientation of
the baseline corresponds to $\psi = 0$. Fig.~\ref{fig:effect_ld} shows
that the impact of limb-darkening for three models: linear, quadratic and
cubic. In all cases the difference due to limb darkening is small, and more
prominent around the peaks.  Investiging further
we have found that the effect is relatively more noticeable for the
baseline range $(0.5-0.8)\times10^9$ and $(1.1-1.5)\times 10^9$. It is
to be noted that for the baseline $1.845 \times 10^9$, the squared
visibility is always more for the limb darkened model than the uniform
model. The reverse is the situation for the baseline higher than $1.845
\times 10^9$.

\begin{figure*}
    \centering
\begin{subfigure}{0.50\textwidth}
    \includegraphics[width=\linewidth]{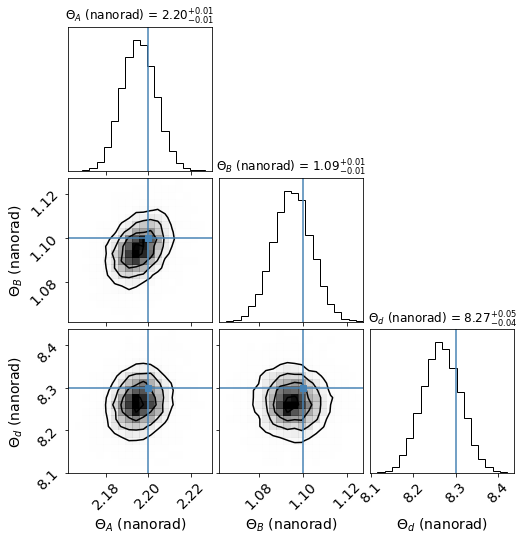}
    \caption{SNR = 370}
    \label{fig:corner_ld_nonlinear}
\end{subfigure}\hfil  
\begin{subfigure}{0.50\textwidth}
    \includegraphics[width=\linewidth]{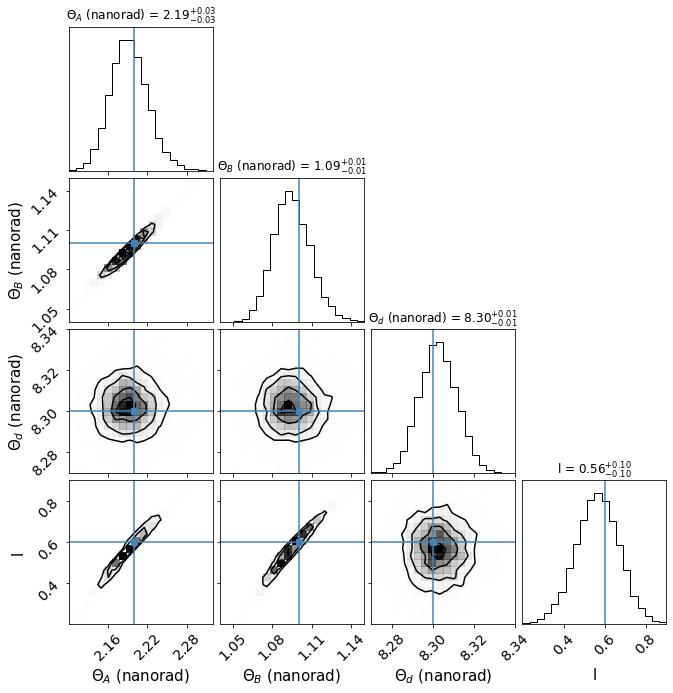}
    \caption{SNR = 1850}
    \label{fig:corner_ld_li}
\end{subfigure}\hfil  
\begin{subfigure}{0.50\textwidth}
    \includegraphics[width=\linewidth]{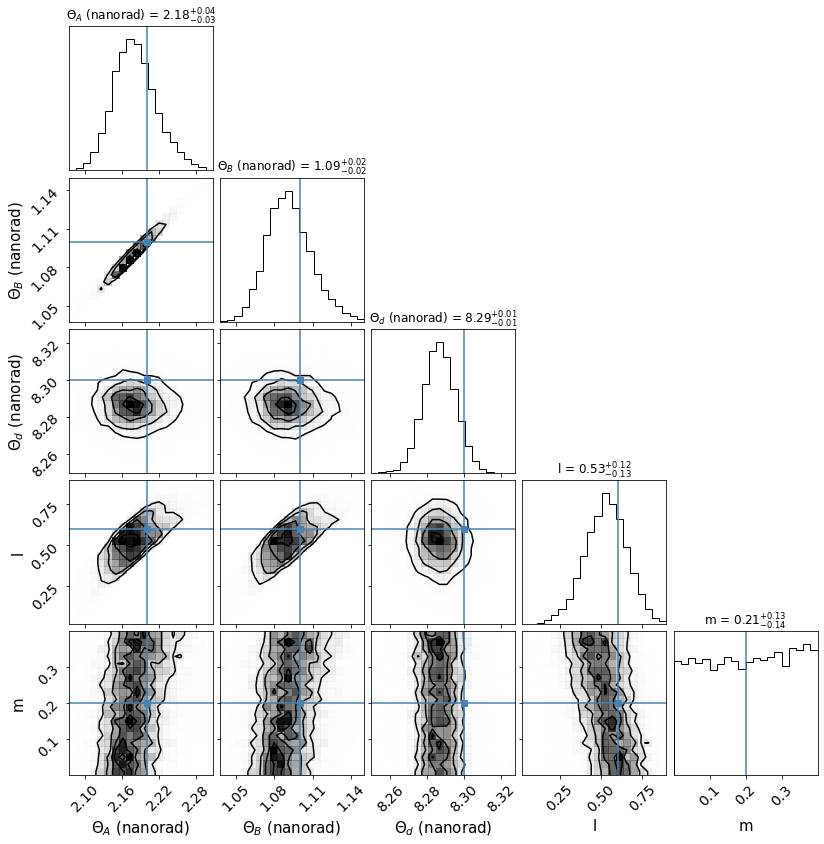}
    \caption{SNR = 1850}
    \label{fig:corner_ld_qd}
\end{subfigure}\hfil  
\begin{subfigure}{0.50\textwidth}
    \includegraphics[width=\linewidth]{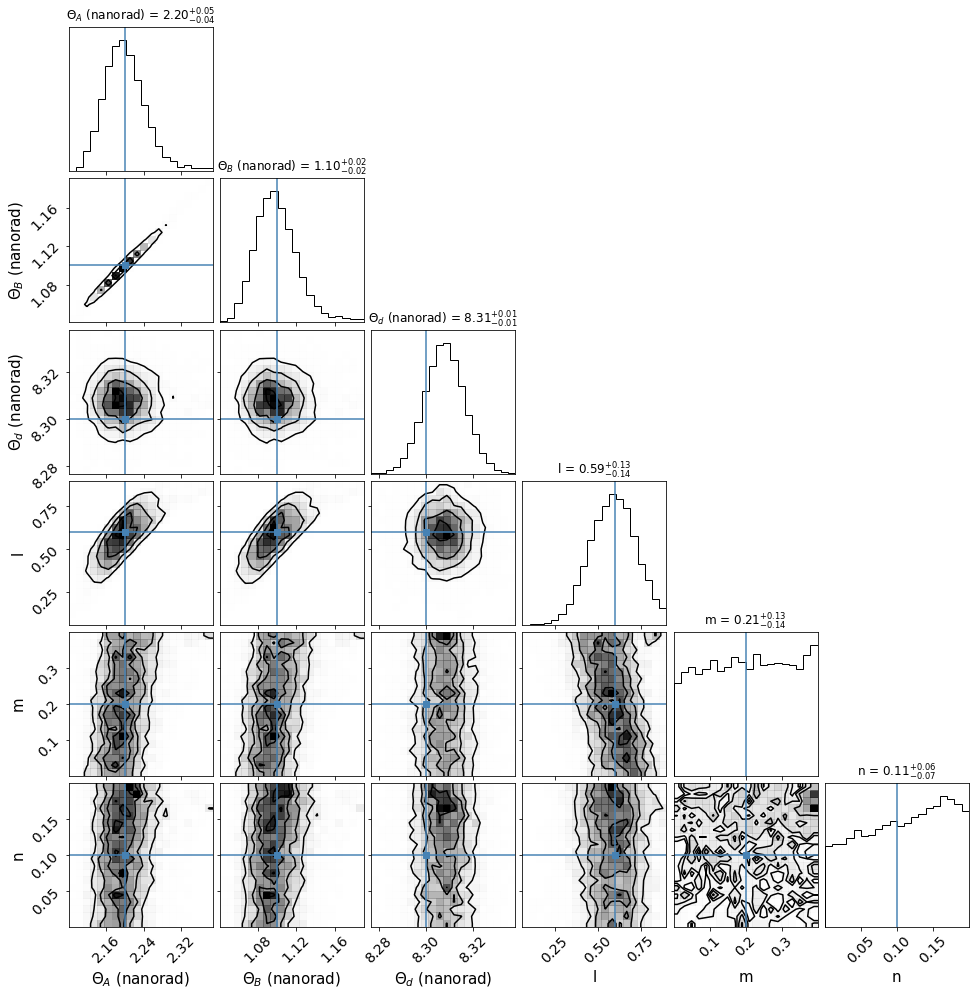}
    \caption{SNR = 1850}
    \label{fig:corner_ld_nl}
\end{subfigure}\hfil  
\caption{Estimation of various numbers of parameters.  The upper
left panel estimates only the angular radii and the angular separation of the stars, with limb-darkening coefficients $l=0.6, m=0.2, n=0.1$ given.  In the upper-right panel, $l$ is estimated while $m$ and $n$ are given. The lower-left corner plot estimates all parameters except $n$ and the
lower right corner plot is for all parameter being estimated. The estimation of $m$ and $n$ is not successful because of their small contribution.}
    \label{fig:corner_ld_all}
\end{figure*}

Fig.~\ref{fig:corner_ld_nonlinear} shows the measurement of the
parameters of the stars with limb-darkening taken into account. The
input values of the limb-darkening coefficients used to generate the
plots are $l=0.6, m=0.2$, and $n=0.1$ and signal-to-noise-ratio equal
to $370$. For estimating the parameters, angular radii of both stars,
and angular separation between them, limb coefficients are taken as
constant.  From Fig.~\ref{fig:effect_ld} and
Fig.~\ref{fig:corner_ld_all} it is clear that the effect of
limb-darkening of stars can be measurable and estimation of
limb-coefficient is also possible.  Fig.~\ref{fig:corner_ld_all} shows
the measurement of all the limb-darkening coefficients together with
the parameters of the stars. For the linear model of the
limb-darkening of stars, we have good recovery of the limb darkening
coefficients $l$. On the other hand, for complex limb darkening models
the recovery of the nonlinear limb-darkening coefficients $m$ and $n$
is not unsuccessful (Fig.~\ref{fig:corner_ld_qd} and
Fig.~\ref{fig:corner_ld_nl}). However, in all cases, we have managed
to recover both angular radii and angular separation between stars
also linear limb-darkening coefficient with good accuracy.

\section{Discussion}

Recent years have seen a resurgence of interest in intensity
interferometry, and many science cases have been discussed in the
literature \citep{2019BAAS...51c.275K}.  In this paper we argue that
radius measurements of binary stars are an interesting science case
for intensity-interferometry observatories now in development.  The
basic idea was known during the earlier generation of intensity
interferometry: \cite{1971MNRAS.151..161H} observed Spica using
intensity interferometry; combining the interferometric observations
with spectroscopy, they reconstructed the binary orbit and inferred
the distance to the system, weighed both the stars, and measured the
radius of the brighter star; the radius of the second star was the
main parameter that could not be determined. It is plausible that the
signal-to-noise needed to measure the radius of the second star was
available even in 1971, but interferometric image reconstruction was
still in its infancy. The interferometric pattern of a binary looks
quite similar to a single star modulated with fringes
(equ.~\eqref{equ:squared_visibility}) and significant computing is
needed to disentangle the contributions of two stars.

Nowadays, several image-reconstruction algorithms and codes are
available \citep{2012A&ARv..20...53B} including developments specific
to intensity interferometry \citep{2015MNRAS.453.1999N}.  In
particular, the reconstruction of an image of two stars from simulated
data has been demonstrated \citep[e.g., Figures 4--6
  in][]{DRAVINS2013331}.  In this work, however, we suggest that
rather than a general-purpose reconstruction of an image of two stars,
parameter-fitting of the observable directed towards inferring the
system parameters may be useful.  Parameters that are not of interest
but are unavoidable, such as the normalisation of the visibility, can
be marginalised out.  Markov chain Monte Carlo is well suited to this
task, and efficient implementations are readily available.  We used
the {\tt emcee} package.

In this paper we carry out a number of simulations, showing that with
sufficient SNR, it is possible to extract both stellar radii, the
angular separation between the stars, and linear limb-darkening.  The
SNR of the simulations can be related to source properties, telescope
parameters, and observing time (eqns.~\ref{equ:snr} and
\ref{equ:specAB}) using well-known expressions for signal and noise in
intensity interferometry.  For example, suppose the spectral
brightness times the effective area (eqns.~\ref{equ:snr} and
\ref{equ:specAB}) is $10^{-4}\rm\,s^{-1}\,Hz^{-1}$.  This will be the
case for zero-magnitude source and an effective collecting around of
about $1\rm\,m^2$, or a 5~mag source and an effective collecting
area of about $100\rm\,m^2$, and so on.  With
nanosec time resolution and an observing time of $10^5\rm\,s$ (2--3
nights) the total SNR will be $\sim1000$.  The
  total observing time will of course be distributed among many
  different baselines.  Thus, there could be 100 baselines obtained
  through pairs of light buckets and the Earth's rotation, each
  observed for $10^3\rm\,s$ for SNR contribution of 100, added
  according to eqn.~\eqref{equ:SNR_total}.

We consider two possible approaches to parameter-fitting for a
binary-star system.

If the stars have different effective temperatures and hence
differently-shaped spectra, and are observed at least two wavelengths,
it is possible to separate the visibility contributions of each stars.
There are several complications to the idea.  The observable has to be
square-rooted while taking into account possible sign changes.
Moreover, the observations would need to be perpendicular to the line
joining the stars, where the visibility is real.  Then, visibility
contributions have to be compared at the same values of dimensionless
baseline at two different wavelengths, which will correspond to
different lengths of the physical baseline.  Finally, the visibility
separation tends to amplify the noise.  Nevertheless, the simulations
indicate that measurement of both stellar radii is possible by this method.

Our second approach is multi-parameter fitting to the observed HBT
correlation (squared visibility).  Provided the parametric form being
fitted is correct, one expects higher accuracy in the parameters.  We
consider a model with two discs in the sky with limb-darkening up to
third order, at some separation from each other.  Rough prior
estimates on the radii of the two discs and their separation are
assumed.  A limb-darkening model with linear, quadratic, and cubic
terms is used.  If the limb-darkening parameters are assumed known, a
total SNR of 350 suffices to estimate the angular radii of the stars
and separation between the stars with reasonable accuracy.  A higher
total SNR of about 1800 is required to estimate the linear
limb-darkening coefficient.  The higher-order coefficients could not
be recovered at this SNR, but that fact does not degrade the estimates
of the other parameters.

The precision of parameter recovery improves with increasing
total SNR, but it also has a strong dependence on
the sampling of the $(u,v)$ plane.  That is to say, even if two
regions of the $(u,v)$ plane may contribute equally to the total SNR
for given observing time, one of the regions may be much more useful
for parameter estimation.  We suggest that the derivatives of the
visibility with respect to the parameters can help identify the most
important parts of the interferometric plane.  By concentrating on
parts of $(u,v)$ plane so identified, we found that a total SNR of 30
was sufficient to recover the angular radii and separation.

These results indicate that binary stars are of
  considerable interest for current and future programs in
  intensity-interferometry.  Individual targets, however, will pose
  their own special challenges. Three questions in particular need
  further study.  One is how to allow for the finite size of the light buckets, which we have not considered at all in this work, but will certainly be an issue for wide binaries where the cosine modulation in the visibility can vary within the diameter of light collector.
A second question that will arise is
whether the $(u,v)$ sampling available from a
  particular observatory is sufficient to recover all the desired
  parameters.  The pioneering work of \cite{1971MNRAS.151..161H} had
  the advantage that the light buckets were moveable on a circular
  railway track, making any desired $(u,v)$ up to a maximum radius
  possible.  Current and planned intensity interferometers are
  restricted to what the rotation of the Earth provides.  That said,
  the observable (eqn.~\ref{equ:squared_visibility}) does not have an
  especially complicated dependence on $(u,v)$ --- it is circular with
  a sinusoidal modulation along one direction --- which suggests that
  a dense sampling of the interferometric plane is not essential for
  recovering the stellar radii and separation.  The simulations shown
  in Figs.~\ref{fig:HBT_corr_msk} and \ref{fig:corner_method2_msk}
  further indicate that even if a region of the $(u,v)$ plane is
  unobserved, the parameters may nonetheless be recovered, because the
  functional form of the observable is known.  This brings us to the third question needing further study: how robust is the parameter
  recovery to errors in the parameterisation itself?  Gross errors
  like mistaking a three-star system for a binary are unlikely,
  because spectroscopy would provide an independent check of the
  binary nature of a system.  But a poor limb-darkening model, or
  unmodelled rotational flattening or star-spots, are of some concern.
  Here the results shown in Fig.~\ref{fig:corner_ld_all} are of
  interest.  They show that parameters (in that case higher-order
  limb-darking parameters) which are not constrained at the available
  SNR get automatically marginalised out by the MCMC, and do not bias
  those parameters which \textit{can} be inferred.  This argues for
  retaining possibly-relevant parameters in a model even if they
  cannot be constrained but only marginalised over.

\section*{Data Availability}
The simulated data and code to generate all the corner plots are included in the supplementary material.

\bibliographystyle{mnras}

\bibliography{main.bib}

\end{document}